# Stock price reaction to power outages following extreme weather events: Evidence from Texas power outage

**Sherry Hu**

Liberty High School in Frisco, Texas

sherryshu2005@gmail.com

**Kose John**

New York University

kjohn@nyu.stern.edu

**Balbinder Singh Gill**

Stevens Institute of Technology

bgill@stevens.edu

October 29, 2022

# Stock price reaction to power outages following extreme weather events: Evidence from the Texas power outage

**Abstract**

In this study, we evaluate the effects of natural disasters on the stock (market) values of firms located in the affected counties. We are able to measure the change in stock prices of the firms affected by the 2021 Texas winter storm. To measure the abnormal return due to the storm, we use four different benchmark models: (1) the market-adjusted model, (2) the market model, (3) the Fama-French three-factor model, and (4) the Fama French plus momentum model. These statistical models in finance characterize the normal risk-return trade-off.

# 1. Introduction

Winter Storm Uri's damage caused the worst energy infrastructure failure in the history of Texas, leaving residents with severe water, food, and heat shortages. Altogether, over 4.5 million homes and businesses were left without power (Pollock, 2021), some for several days, and at least 246 people were killed directly or indirectly because of the crisis (TDSHS, 2021). As a result of the deadly storm, experts estimate around "$130 billion in damages and economic losses in Texas" (Puleo, 2021) while "estimates of insured losses … range from $10 billion to $20 billion" (Golding, et al., 2021).

## 2. Historical Overview of Texas Power Grid

The issue of Texas' deregulated power grid has been a hotly debated issue for years. Recently, due to extensive news coverage of the devastating Valentine's Day winter storms and the accompanying blackouts that swept across every Texas county, leaving millions of residents at its peak without power, a national spotlight has been shone on the topic once again, leaving many to scrutinize the actual effectiveness and reliability of a standalone power grid.

The isolation of the Texas grid from the rest of the United States can be traced back to the Federal Power Act passed by Congress in 1935 which essentially established/delegated the federal government as the sole jurisdiction over interstate electrical transmission through the Federal Power Commission (FPC), now the Federal Energy Regulatory Commission (FERC). However, the Texas Interconnected System, made up of connected city grids, wanted to avoid federal regulation. Through isolating their properties to within the state, "Texas utilities avoided being subjected to federal rules" (Gailbraith, 2011) and interstate electric commerce.

### 2.1. Formation of Electric Reliability Council of Texas (ERCOT)

In 1965, a significant blackout affecting 30 million in the Northeastern United States and surrounding areas led the electricity industry to form a "council of regional electricity coordination organizations" (Lin, 2021) called the National Electric Reliability Council in 1968, now known as the North American Electric Reliability Corporation (NERC) with a broader reach across North America.

In response to NERC, the Electric Reliability Council of Texas (ERCOT) was formed in 1970 to manage the Texas grid, regulated by the governor-appointed Public Utility Commission to oversee "the price of power" and balance "supply and demand in the grid" (Cameron, et al., 2021). Despite a few instances of sharing power with other regions, the Texas grid has stayed relatively independent since the establishment of ERCOT.

Fast-forward to present-day, with Texas maintaining a "consistent load growth in recent decades due to [its] strong economy and increasing population," ERCOT now manages an electric grid including "567 electricity generators, 127 retail electric providers, and 85 transmission companies" (Cameron, et al., 2021) that serves approximately 26 million Texas residents, or around "90% of the electric load in Texas" (Lin, 2021). With so much of the state grid under its control, ERCOT has received considerable disapproval for its lack of reliability during the events of February 2021. Residents and onlookers alike criticized how ERCOT was irresponsible enough to overlook the possibility of an extreme weather event that carried critical consequences. Others blamed how the Texas government and energy companies had held off on winterizing the power plants, for example, by adding insulation to pipes and dismantling the stand-alone power grid to rely on power from other regions, the state would not have been as severely affected as it was.

### 2.2. Winter Storm Uri

On February 13th, 2021, an extreme cold weather event swept over Texas and its surrounding regions, leading ERCOT to issue an Emergency Notice for the weather. Texas, a state notorious for its hot weather and humid climate, was unequipped for such conditions, unlike colder Northern states. "The ability to restore more power [was] contingent on more generation coming back online" which was not possible due to "a range of factors including frozen wind turbines, low gas pressure and frozen instrumentation" (Booker and Romo).

While February 13th was the "first day that large generators began to unexpectedly go offline", it was on February 14th when the electricity load was "approaching available generation" and ERCOT urged consumers to conserve energy usage and issued "multiple watches regarding power supply shortages" (Lin, 2021). Later, just minutes after midnight on Monday, February 15, ERCOT declared an Energy Emergency Alert Level 1.

By 9:00 a.m. on February 15th, the total complete and partial unavailability of the power supply increased to over 50,000 megawatts (approximately 40% of the total maximum output in ERCOT). Researchers at the University of Texas at Austin's Energy Institute (2021, p. 29) would later note that "levels of outages and derates would change over the event, but would not return to pre-blackout levels until the afternoon of February 19".

We discover that stock listed businesses in the counties affected by the power outage after the extreme event experienced a sizable decline in market value. This outcome suggests that shareholders of these companies are eager to cut their losses by selling their shares as soon as the power outage occurs. This is due to their expectation that this power outage will increase the level of uncertainty surrounding how their company will manage the negative effects of such a large-scale power outage. It demonstrates how unprepared these businesses were for power outages of this magnitude.

The remainder of this paper is organized as follows. Section 2 describes our sample and data composition. Section 3 discusses our empirical results. Section 4 concludes our paper.

## 3. Sample and data composition

We select companies in counties that were affected by the power outage on Feb 16, 2021, and Feb 17, 2021 respectively. We use power outage data from Miller (2021) to identify the counties that were affected by the power outage. These companies are publicly traded companies. Financial data of these companies are provided by the Compustat Database. Table 1 gives an overview of summary statistics (number of observations, mean, median, standard deviation, min, and max) of the most important characteristics of these firms such as the firm size, market value of the firm, and profitability of the firm.

Table 1 presents the overview of the firms hit by the power outage. Table 2 shows the descriptive statistics of the sample. The number of observations across the characteristics of the firms in the sample varies due to data availability. We observe that we have a mix of large firms in terms of size and market value, profitable firms, and firms with different financial structures.

**Insert Tables 1 and 2**

## 4. Empirical results

We use the U.S. Daily Event Study tool that is provided by Wharton Research Data Services[1]. The length of the time in trading days used to estimate the expected return and residual return variance is 10. The minimum number of non-missing return observations within the estimation window required to produce estimates of expected return is 7. The number of gap days is 10. The event window starts two trading days before the event occurrence and ends two days after the event occurrence. We use four asset pricing models to calculate the expected returns. These four asset pricing models are the Market-Adjusted Model, Market Model, Fama-French Three Factor Model, and Fama-French Plus momentum Model. The U.S. Daily Event Study tool uses daily stock data from the Center for Research in Security Prices (CRSP). The U.S. Daily Event Study tool shows graphically the results of the event study.

Figure 1 shows the results of the event study of the effect of the power outage on the cumulative abnormal returns by estimated model of the expected return. The market model (base model) shows that the power outage had on average a negative effect of the stock return of companies that were affected by this power outage. The mean CAR is statistically significantly declining following the power outage event. We observe a similar pattern for the remaining three types of market model. As a result, this result is robust for the type of asset pricing model used to calculate the expected returns.

[1] Event study is a standard empirical method used in finance to evaluate the abnormal changes in the stock market value of companies affected by an unexpected event. The idea is to estimate the abnormal changes in stock price during the event days. The normal return that corresponds to the systematic risk of the stock is called the normal return. We use four common models to estimate the normal bench-mark model.

**Insert Figures 1, 2,3, and 4 here**

## 5. Conclusion

This paper investigates the effect of a power outage caused by an extreme weather event on the market value of firms located in counties that were affected by this power outage. By using event study methodology, we find that the market value of power affected firms statistically significantly decreased. Investors expect that power outage affected firms will face uncertain future ex-ante the power outage.

**Table 1: Sample overview of firms hit by the power outage**

This Table provides an overview of firms affected by the power outage following the extreme weather events. These firms are all stock listed firms.

| CUSIP | Company Name | Postal Code | Standard Industry Classification Code | State |
|---|---|---|---|---|
| 02376R102 | AMERICAN AIRLINES GROUP INC | 76155 | 4512 | TX |
| 125141101 | CECO ENVIRONMENTAL CORP | 75254 | 3564 | TX |
| 006351308 | ADAMS RESOURCES & ENERGY INC | 77027 | 5172 | TX |
| 049904105 | ATRION CORP | 75002-4211 | 3841 | TX |
| 023771009 | AMERICAN AIRLINES INC | 76155 | 4512 | TX |
| 02772A109 | AMERICAN NATIONAL GROUP INC | 77550-7999 | 6311 | TX |
| 02772A109 | AMERICAN NATIONAL GROUP INC | 77550-7999 | 6311 | TX |
| 029174109 | AMERICAN REALTY INVESTORS | 75234 | 6512 | TX |
| 029174109 | AMERICAN REALTY INVESTORS | 75234 | 6512 | TX |
| 03743Q108 | APA CORP | 77056-4000 | 1311 | TX |
| 894648104 | TRECORA RESOURCES | 77478 | 2911 | TX |
| 002474104 | AZZ INC | 76107 | 3640 | TX |
| 08499Z004 | BURLINGTON NORTHERN SANTA FE | 76131-2830 | 4011 | TX |
| 05946K002 | BBVA COMPASS BANCSHARES INC | 77056 | 6020 | TX |
| 05946K002 | BBVA COMPASS BANCSHARES INC | 77056 | 6020 | TX |
| 109641100 | BRINKER INTL INC | 75019 | 5812 | TX |
| 45174J509 | IHEARTMEDIA INC | 78258 | 4832 | TX |
| 200340107 | COMERICA INC | 75201 | 6020 | TX |
| 200340107 | COMERICA INC | 75201 | 6020 | TX |
| 201723103 | COMMERCIAL METALS | 75039 | 3312 | TX |
| 754198109 | RAVE RESTAURANT GROUP INC | 75056 | 6794 | TX |
| 893617209 | TRANSCONTINENTAL RLTY INVS | 75234 | 6512 | TX |
| 893617209 | TRANSCONTINENTAL RLTY INVS | 75234 | 6512 | TX |
| 08986R309 | BIGLARI HOLDINGS INC | 78257 | 5812 | TX |
| 229899109 | CULLEN/FROST BANKERS INC | 78205 | 6020 | TX |
| 229899109 | CULLEN/FROST BANKERS INC | 78205 | 6020 | TX |
| 239360100 | DAWSON GEOPHYSICAL CO | 79701 | 1382 | TX |
| 254543101 | DIODES INC | 75024 | 3674 | TX |
| 25820R105 | DORCHESTER MINERALS -LP | 75219-4541 | 6792 | TX |
| 34354P105 | FLOWSERVE CORP | 75039 | 3561 | TX |
| G48833118 | WEATHERFORD INTL PLC | 77056 | 1381 | TX |
| 049560105 | ATMOS ENERGY CORP | 75240 | 4924 | TX |
| 293389102 | ENNIS INC | 76065 | 2761 | TX |
| 30231G102 | EXXON MOBIL CORP | 75039-2298 | 2911 | TX |
| 307675108 | FARMER BROTHERS CO | 76262 | 2090 | TX |
| 343412102 | FLUOR CORP | 75039 | 1600 | TX |
| 358435105 | FRIEDMAN INDUSTRIES INC | 75601 | 3310 | TX |
| 60935Y208 | MONEYGRAM INTERNATIONAL INC | 75201 | 6099 | TX |
| 60935Y208 | MONEYGRAM INTERNATIONAL INC | 75201 | 6099 | TX |
| 406216101 | HALLIBURTON CO | 77032 | 1389 | TX |
| 859241101 | STERLING INFRASTRUCTURE INC | 77380 | 1600 | TX |
| G4388N106 | HELEN OF TROY LTD | 79912 | 3634 | TX |
| 15189T107 | CENTERPOINT ENERGY INC | 77002 | 4931 | TX |
| 49456B101 | KINDER MORGAN INC | 77002 | 4923 | TX |
| 74158E104 | PRIMEENERGY RESOURCES CORP | 77024 | 1311 | TX |
| 494368103 | KIMBERLY-CLARK CORP | 75261-9100 | 2621 | TX |
| 497266106 | KIRBY CORP | 77007 | 4400 | TX |
| 75281A109 | RANGE RESOURCES CORP | 76102 | 1311 | TX |
| 565849106 | MARATHON OIL CORP | 77024-2217 | 1311 | TX |
| 568423107 | MARINE PETROLEUM TRUST | 75219 | 6792 | GA |
| 58155Q103 | MCKESSON CORP | 75039 | 5122 | TX |
| 590660106 | MESA ROYALTY TRUST | 77002 | 6792 | TX |
| 592770101 | MEXCO ENERGY CORP | 79701 | 1311 | TX |
| 59408Q106 | MICHAELS COS INC | 75063 | 5945 | TX |
| 595135104 | MICROPAC INDUSTRIES INC | 75040 | 3674 | TX |
| 626717102 | MURPHY OIL CORP | 77024 | 1311 | TX |
| 629156407 | NL INDUSTRIES | 75240-2620 | 3562 | TX |
| 88033G407 | TENET HEALTHCARE CORP | 75254 | 8062 | TX |
| 651718504 | NEWPARK RESOURCES | 77381 | 5160 | TX |
| 674599105 | OCCIDENTAL PETROLEUM CORP | 77046-0521 | 1311 | TX |
| 675232102 | OCEANEERING INTERNATIONAL | 77041 | 1389 | TX |
| 11040G103 | BRISTOW GROUP INC | 77042 | 4522 | TX |
| 79471V105 | SALIENT MIDSTREAM & MLP FUND | 77027 | 6726 | TX |
| 698465002 | PANHANDLE EASTERN PIPE LINE | 75225 | 4922 | TX |
| 382410843 | GOODRICH PETROLEUM CORP | 77002 | 1311 | TX |
| 70788V102 | RANGER OIL CORPORATION | 77084 | 1311 | TX |
| 714236106 | PERMIAN BASIN ROYALTY TRUST | 75219 | 6792 | TX |
| 20825C104 | CONOCOPHILLIPS | 77079-2703 | 1311 | TX |
| 911922102 | U S LIME & MINERALS | 75240 | 1400 | TX |
| 785688102 | SABINE ROYALTY TRUST | 75219 | 6792 | TX |
| 798241105 | SAN JUAN BASIN ROYALTY TR | 77056 | 6792 | TX |
| 806857108 | SCHLUMBERGER LTD | 77056 | 1389 | TX |
| 25400W102 | DIGITAL TURBINE INC | 78701 | 7372 | TX |
| 817565104 | SERVICE CORP INTERNATIONAL | 77019 | 7200 | TX |
| 844741108 | SOUTHWEST AIRLINES | 75235-1611 | 4512 | TX |
| 00206R102 | AT&T INC | 75202 | 4812 | TX |
| 845467109 | SOUTHWESTERN ENERGY CO | 77389 | 4923 | TX |
| 845743004 | SOUTHWESTERN PUBLIC SVC CO | 79101 | 4911 | TX |
| 860372101 | STEWART INFORMATION SERVICES | 77056 | 6361 | TX |
| 860372101 | STEWART INFORMATION SERVICES | 77056 | 6361 | TX |
| 82836G102 | SILVERBOW RESOURCES INC | 77024 | 1311 | TX |
| 871829107 | SYSCO CORP | 77077-2099 | 5140 | TX |
| 878155100 | TEAM INC | 77478 | 8734 | TX |
| 882508104 | TEXAS INSTRUMENTS INC | 75243 | 3674 | TX |
| 882587009 | TEXAS NEW MEXICO POWER CO | 75067 | 4911 | TX |
| 88262P102 | TEXAS PACIFIC LAND CORP | 75201 | 6792 | TX |
| 88642R109 | TIDEWATER INC | 77024 | 4400 | TX |
| 83001A102 | SIX FLAGS ENTERTAINMENT CORP | 76011 | 7996 | TX |
| 37959E102 | GLOBE LIFE INC | 75070 | 6311 | TX |
| 37959E102 | GLOBE LIFE INC | 75070 | 6311 | TX |
| 893570002 | TRANSCONTINENTAL GAS PIPE LN | 77056 | 4922 | TX |
| 896522109 | TRINITY INDUSTRIES INC | 75254-2957 | 3743 | TX |
| 928661107 | VOLITIONRX LTD | 78738 | 2835 | TX |
| 902252105 | TYLER TECHNOLOGIES INC | 75024 | 7373 | TX |
| 905581005 | UNION CARBIDE CORP | 77983 | 2860 | TX |
| 911805307 | US ENERGY CORP | 77057 | 1311 | TX |
| 911805307 | US ENERGY CORP | 77057 | 1311 | TX |
| 948741103 | WEINGARTEN REALTY INVST | 77008 | 6798 | TX |
| 948741103 | WEINGARTEN REALTY INVST | 77008 | 6798 | TX |
| 643611106 | NEW CONCEPT ENERGY INC | 75234 | 6512 | TX |
| 89904V101 | TUESDAY MORNING CORP | 75240-6321 | 5331 | TX |
| 29402E102 | ENVELA CORP | 75038-5202 | 5990 | TX |
| 68389X105 | ORACLE CORP | 78741 | 7370 | TX |
| 121899009 | BNSF RAILWAY CO | 76131-2830 | 4011 | TX |
| 416196202 | HARTE HANKS INC | 01824 | 7331 | MA |
| 918905209 | VALHI INC | 75240-2620 | 2810 | TX |
| 86765K109 | SUNOCO LP | 75225 | 5172 | TX |
| 452926108 | INCOME OPPORTUNITY RLTY INVS | 75234 | 6162 | TX |
| 14067E506 | CAPSTEAD MORTGAGE CORP | 75225-4404 | 6798 | TX |
| 14067E506 | CAPSTEAD MORTGAGE CORP | 75225-4404 | 6798 | TX |
| 08579X101 | BERRY CORP | 75248 | 1311 | TX |
| 70532Y303 | PEDEVCO CORP | 77079 | 1311 | TX |
| 866142409 | SUMMIT MIDSTREAM PARTNERS LP | 77002 | 1311 | TX |
| 62482R107 | MR COOPER GROUP INC | 75019 | 6162 | TX |
| 62482R107 | MR COOPER GROUP INC | 75019 | 6162 | TX |
| 808513105 | SCHWAB (CHARLES) CORP | 76262 | 6282 | TX |
| 808513105 | SCHWAB (CHARLES) CORP | 76262 | 6282 | TX |
| G9325C105 | VANTAGE DRILLING INTL | 77056 | 1381 | TX |
| 723787107 | PIONEER NATURAL RESOURCES CO | 75039 | 1311 | TX |
| 94106L109 | WASTE MANAGEMENT INC | 77002 | 4953 | TX |
| 24703L202 | DELL TECHNOLOGIES INC | 78682 | 3571 | TX |
| 231647207 | CUSHING NEXTGEN INFRSTR INC | 75201 | 6726 | TX |
| 13123X508 | CALLON PETROLEUM CO/DE | 77042 | 1311 | TX |
| 892332008 | TOYOTA MOTOR CREDIT CORP | 75024 | 6141 | TX |
| 91913Y100 | VALERO ENERGY CORP | 78249 | 2911 | TX |
| 055630107 | BP PRUDHOE BAY ROYALTY TRUST | 77002 | 6792 | TX |
| 05565A202 | BNP PARIBAS | 75009 | 6020 | |
| 05565A202 | BNP PARIBAS | 75009 | 6020 | |
| 83364L109 | SOCIETE GENERALE GROUP | 75009 | 6020 | |
| 83364L109 | SOCIETE GENERALE GROUP | 75009 | 6020 | |
| 74836W203 | QUEST RESOURCE HOLDING CORP | 75056 | 4950 | TX |
| 23283R100 | CYRUSONE INC | 75201 | 6798 | TX |
| 858568108 | STELLUS CAPITAL INVESTMENT | 77027 | 6797 | TX |
| 958669103 | WESTERN MIDSTRM PRTNRS LP | 77380 | 4922 | TX |
| 172755100 | CIRRUS LOGIC INC | 78701 | 3674 | TX |
| 41043F208 | HANGER INC | 78758-7807 | 8093 | TX |
| 26875P101 | EOG RESOURCES INC | 77002 | 1311 | TX |
| 98974P100 | ZIX CORP | 75204-2960 | 7370 | TX |
| 174740100 | CITIZENS INC | 78758 | 6311 | TX |
| 174740100 | CITIZENS INC | 78758 | 6311 | TX |
| 28413L105 | ELAH HOLDINGS INC | 75205 | 9995 | TX |
| 28413L105 | ELAH HOLDINGS INC | 75205 | 9995 | TX |
| 638517102 | NATIONAL WESTERN LIFE GROUP | 78759-5415 | 6311 | TX |
| 638517102 | NATIONAL WESTERN LIFE GROUP | 78759-5415 | 6311 | TX |
| 37045V001 | GENERAL MOTORS FINL CO INC | 76102 | 6141 | TX |
| 37045V001 | GENERAL MOTORS FINL CO INC | 76102 | 6141 | TX |
| 87484T108 | TALOS ENERGY INC | 77002 | 1311 | TX |
| 45384B106 | INDEPENDENT BK GRP INC | 75070-1711 | 6020 | TX |
| 45384B106 | INDEPENDENT BK GRP INC | 75070-1711 | 6020 | TX |
| 23636T100 | DANONE SA | 75009 | 2000 | |
| 718549207 | PHILLIPS 66 PARTNERS LP | 77042 | 4610 | TX |
| 80283M101 | SANTANDER CONSUMER USA HLDGS | 75201 | 6141 | TX |
| 72651A207 | PLAINS GP HOLDINGS LP | 77002 | 5171 | TX |
| 451622203 | IDEAL POWER INC | 78735 | 3613 | TX |
| 03823U102 | APPLIED OPTOELECTRONICS INC | 77478 | 3674 | TX |
| 50187T106 | LGI HOMES INC | 77380 | 1531 | TX |
| 226718104 | CRITEO SA | 75009 | 7311 | |
| 210751103 | CONTAINER STORE GROUP | 75019-3863 | 5700 | TX |
| 10482B101 | BRAEMAR HOTELS & RESORTS INC | 75254 | 6798 | TX |
| 78573M104 | SABRE CORP | 76092 | 7370 | TX |
| 76118L102 | RESONANT INC | 78759 | 3674 | TX |
| 21872L104 | COREPOINT LODGING INC | 75062 | 6798 | TX |
| 03890D108 | ARAVIVE INC | 77098 | 2836 | TX |
| 178587101 | CITY OFFICE REIT INC | 75201 | 6798 | TX |
| 74736L109 | Q2 HOLDINGS INC | 78729 | 7370 | TX |
| 29336T100 | ENLINK MIDSTREAM LLC | 75201 | 4922 | TX |
| 125525584 | CREATIVE MEDIA & COM TR CORP | 75252 | 6798 | TX |
| 125525584 | CREATIVE MEDIA & COM TR CORP | 75252 | 6798 | TX |
| 17878Y207 | CIVEO CORP | 77002 | 7011 | TX |
| 140501107 | CAPITAL SOUTHWEST CORP | 75225 | 6797 | TX |
| 140501107 | CAPITAL SOUTHWEST CORP | 75225 | 6797 | TX |
| 05601C105 | BGSF INC | 75024 | 7363 | TX |
| 05601C105 | BGSF INC | 75024 | 7363 | TX |
| 67011P100 | NOW INC | 77041 | 5084 | TX |
| 960417103 | WESTLAKE CHEMICAL PRTNRS LP | 77056 | 2860 | TX |
| 92763M105 | VIPER ENERGY PARTNERS LP | 79701 | 6792 | TX |
| 127097103 | COTERRA ENERGY INC | 77024 | 1311 | TX |
| 24790A101 | DENBURY INC | 75024 | 1311 | TX |
| 32026V104 | FIRST FOUNDATION INC | 75201 | 6020 | TX |
| 32026V104 | FIRST FOUNDATION INC | 75201 | 6020 | TX |
| 92556D106 | VIA RENEWABLES INC | 77079 | 4931 | TX |
| 822634101 | SHELL MIDSTREAM PARTNERS LP | 77079 | 4610 | TX |
| 781386305 | RUMBLEON INC | 75038 | 5961 | TX |
| 453415606 | INDEPENDENCE CONTRACT DRLLNG | 77070 | 1381 | TX |
| G7041T139 | PETRO-VICTORY ENERGY CORP | 76106 | 1311 | TX |
| 88162F105 | TETRA TECHNOLOGIES INC/DE | 77380 | 1389 | TX |
| 903318103 | USD PARTNERS LP | 77002 | 4610 | TX |
| 923451108 | VERITEX HOLDINGS INC | 75225 | 6020 | TX |
| 923451108 | VERITEX HOLDINGS INC | 75225 | 6020 | TX |
| 92645B103 | VICTORY CPTL HLDGS INC | 78256 | 6282 | TX |
| 89679E300 | TRIUMPH BANCORP INC | 75251 | 6036 | TX |
| 89679E300 | TRIUMPH BANCORP INC | 75251 | 6036 | TX |
| 428103105 | HESS MIDSTREAM LP | 77010 | 1311 | TX |
| 24372A305 | KOIL ENERGY SOLUTIONS INC | 77049 | 1382 | TX |
| 91544A109 | UPLAND SOFTWARE INC | 78701-3788 | 7370 | TX |
| 392709101 | GREEN BRICK PARTNERS INC | 75093 | 1531 | TX |
| 21075N204 | CONTANGO OIL & GAS CO | 76102 | 1311 | TX |
| 459044103 | INTL BANCSHARES CORP | 78042-1359 | 6020 | TX |
| 459044103 | INTL BANCSHARES CORP | 78042-1359 | 6020 | TX |
| 044104107 | ASHFORD INC | 75254 | 6282 | TX |
| 044104107 | ASHFORD INC | 75254 | 6282 | TX |
| 079481404 | BELLICUM PHARMACEUTICALS INC | 77098 | 2836 | TX |
| 83191H107 | SMART SAND INC | 77380 | 1400 | TX |
| 92735P103 | VINE ENERGY INC | 75024-6642 | 1311 | TX |
| 123159105 | BURZYNSKI RESEARCH INSTITUTE | 77055 | 2836 | TX |
| 69888T207 | PAR PACIFIC HOLDINGS INC | 77024 | 2911 | TX |
| 27032D304 | EARTHSTONE ENERGY INC | 77380 | 1311 | TX |
| 63902N106 | NATURALSHRIMP INC | 75240 | 0200 | TX |
| 65341D102 | NEXPOINT RESIDENTIAL TR INC | 75201 | 6798 | TX |
| 65341D102 | NEXPOINT RESIDENTIAL TR INC | 75201 | 6798 | TX |
| 34960W106 | FORTERRA INC | 75062 | 3272 | TX |
| 98400H102 | XBIOTECH INC | 78744 | 2836 | TX |
| 09225M101 | BLACK STONE MINERALS LP | 77002 | 1311 | TX |
| 79957L100 | SANARA MEDTECH INC | 76102 | 3842 | TX |
| 26922A842 | U S GLOBAL JETS ETF | 78229 | 6722 | TX |
| 462044207 | ION GEOPHYSICAL CORP | 77042-2855 | 1382 | TX |
| 974155103 | WINGSTOP INC | 75001 | 5812 | TX |
| 30227H106 | EXTERRAN CORP | 77041 | 1389 | TX |
| 003830304 | ABRAXAS PETROLEUM CORP/NV | 78258 | 1311 | TX |
| 848550208 | SPINDLETOP OIL & GAS CO | 75230-1279 | 1311 | TX |
| 221006109 | CORVEL CORP | 76109 | 6411 | TX |
| 23753F107 | DASEKE INC | 75001 | 4213 | TX |
| 00773J103 | AEGLEA BIOTHERAPEUTICS INC | 78746 | 2836 | TX |
| 126402106 | CSW INDUSTRIALS INC | 75240 | 2800 | TX |
| 22757R109 | CROSS TIMBERS ROYALTY TRUST | 75219 | 6792 | TX |
| 04649U102 | ASURE SOFTWARE INC | 78746 | 7370 | TX |
| 46121E205 | INTRUSION INC | 75081 | 7372 | TX |
| 90337L108 | U S PHYSICAL THERAPY INC | 77042 | 8000 | TX |
| 23331A109 | D R HORTON INC | 76011 | 1531 | TX |
| 863167201 | STRATUS PROPERTIES INC | 78701 | 6552 | TX |
| 863167201 | STRATUS PROPERTIES INC | 78701 | 6552 | TX |
| 292562105 | ENCORE WIRE CORP | 75069 | 3357 | TX |
| 294766100 | EQUUS TOTAL RETURN INC | 77002 | 6797 | TX |
| 40425J101 | HMS HOLDINGS CORP | 75038 | 6411 | TX |
| 57667L107 | MATCH GROUP INC | 75231 | 7370 | TX |
| 04650Y100 | AT HOME GROUP INC | 75074 | 5700 | TX |
| 65506L105 | NOBLE MIDSTREAM PARTNERS LP | 77070 | 4610 | TX |
| 922060108 | VANJIA CORP | 77077 | 1531 | TX |
| 28621V101 | ELEVATE CREDIT INC | 76109 | 6141 | TX |
| 09229E204 | BLACKBOXSTOCKS INC | 75240 | 7370 | TX |
| 60855D200 | MOLECULIN BIOTECH INC | 77007 | 2836 | TX |
| 75615P103 | REATA PHARMACEUTICALS INC | 75024 | 2836 | TX |
| 656357100 | NORRIS INDUSTRIES INC | 76086 | 1311 | TX |
| 27888N307 | ECOARK HOLDINGS INC | 78215 | 1311 | TX |
| 91851C201 | VAALCO ENERGY INC | 77042 | 1311 | TX |
| 02073X105 | ALPHA ENERGY INC | 80401 | 1311 | CO |
| 98585X104 | YETI HOLDINGS INC | 78735 | 3949 | TX |
| 867931701 | SUPERCONDUCTOR TECHNOLOGIES | 78738 | 3670 | TX |
| 40624Q203 | HALLMARK FINANCIAL SERVICES | 75240-2345 | 6331 | TX |
| 40624Q203 | HALLMARK FINANCIAL SERVICES | 75240-2345 | 6331 | TX |
| 011311107 | ALAMO GROUP INC | 78155 | 3523 | TX |
| 268211109 | DASAN ZHONE SOLUTIONS INC | 75024 | 3661 | TX |
| 34988V106 | FOSSIL GROUP INC | 75080 | 3873 | TX |
| 343389102 | FLOTEK INDUSTRIES INC | 77064 | 2891 | TX |
| 68752M108 | ORTHOFIX MEDICAL INC | 75056 | 3841 | TX |
| 133131102 | CAMDEN PROPERTY TRUST | 77046 | 6798 | TX |
| 133131102 | CAMDEN PROPERTY TRUST | 77046 | 6798 | TX |
| 87538X105 | TANDY LEATHER FACTORY INC | 76140-1003 | 5990 | TX |
| 32020R109 | FIRST FINL BANKSHARES INC | 79601 | 6020 | TX |
| 32020R109 | FIRST FINL BANKSHARES INC | 79601 | 6020 | TX |
| 703481101 | PATTERSON-UTI ENERGY INC | 77064 | 1381 | TX |
| 16411R208 | CHENIERE ENERGY INC | 77002 | 1311 | TX |
| 65290C105 | NEXTIER OILFIELD SOLUTNS INC | 77042 | 1389 | TX |
| 81617J301 | SELECT ENERGY SERVICES INC | 77027 | 1389 | TX |
| 03767D108 | APOLLO ENDOSURGERY INC | 78746 | 3841 | TX |
| 49435R102 | KIMBELL ROYALTY PARTNERS LP | 76102 | 1311 | TX |
| 217204106 | COPART INC | 75254 | 5010 | TX |
| 46187W107 | INVITATION HOMES INC | 75201 | 6798 | TX |
| 26969P108 | EAGLE MATERIALS INC | 75225 | 3270 | TX |
| 74347M108 | PROPETRO HOLDING CORP | 79701 | 1389 | TX |
| 87968A104 | TELLURIAN INC | 77002 | 1311 | TX |
| 306137209 | FALCONSTOR SOFTWARE INC | 78701 | 7373 | TX |
| 83418M103 | SOLARIS OILFIELD IF INC | 77024 | 7359 | TX |
| 237266101 | DARLING INGREDIENTS INC | 75038 | 2070 | TX |
| 12739A100 | CADENCE BANCORPORATION | 77056 | 6020 | TX |
| 12739A100 | CADENCE BANCORPORATION | 77056 | 6020 | TX |
| 628877201 | NCS MULTISTAGE HLDG INC | 77070 | 3533 | TX |
| 805111101 | SAVARA INC | 78746 | 2834 | TX |
| 602566101 | MIND TECHNOLOGY INC | 77380 | 3829 | TX |
| 559663109 | MAGNOLIA OIL & GAS CORP | 77046 | 1311 | TX |
| 559663109 | MAGNOLIA OIL & GAS CORP | 77046 | 1311 | TX |
| 400764106 | GUARANTY BANCSHARES INC | 75001 | 6020 | TX |
| 400764106 | GUARANTY BANCSHARES INC | 75001 | 6020 | TX |
| G6375R107 | NATIONAL ENERGY SERVICES REU | 77056 | 1389 | TX |
| 78413P101 | SEACOR MARINE HLDGS INC | 77079 | 4412 | TX |
| 78413P101 | SEACOR MARINE HLDGS INC | 77079 | 4412 | TX |
| 67420T206 | OASIS MIDSTREAM PARTNR | 77002 | 4610 | TX |
| 76009N100 | RENT-A-CENTER INC | 75024 | 7350 | TX |
| 65441V101 | NINE ENERGY SERVICE INC | 77019 | 1389 | TX |
| 636518102 | NATIONAL INSTRUMENTS CORP | 78759-3504 | 7372 | TX |
| 75282U104 | RANGER ENERGY SERVICES | 77042 | 1389 | TX |
| 26922A719 | US GLOBAL GD & PRC MT MI ETF | 78229 | 6722 | TX |
| 05722G100 | BAKER HUGHES CO | 77073-5101 | 1389 | TX |
| 65342K105 | NEXTDECADE CORP | 77002 | 4923 | TX |
| 372446104 | GENPREX INC | 78746 | 2836 | TX |
| 608550109 | MOLECULAR TEMPLATES | 78729 | 2836 | TX |
| 0556EL109 | BP MIDSTREAM PARTNERS | 77079 | 4610 | TX |
| 12481V104 | CBTX INC | 77706 | 6020 | TX |
| 12481V104 | CBTX INC | 77706 | 6020 | TX |
| 78781P105 | SAILPOINT TECHNO HLDG | 78726 | 7370 | TX |
| 43010E404 | HIGHLAND INCOME FUND | 75201 | 6722 | TX |
| 127203107 | CACTUS INC | 77024 | 3533 | TX |
| 30283W302 | FTS INTERNATIONAL INC | 76102 | 1389 | TX |
| 36113U101 | FUSE MEDICAL INC | 75080 | 5047 | TX |
| 84861D103 | SPRT OF TXS BNCSR INC | 77301 | 6020 | TX |
| 84861D103 | SPRT OF TXS BNCSR INC | 77301 | 6020 | TX |
| 15872M104 | CHAMPIONX CORP | 77381 | 3533 | TX |
| 38267D109 | GOOSEHEAD INSURANCE | 76262 | 6411 | TX |
| 90290N109 | USA COMPRESSION PRTNRS LP | 78701 | 1389 | TX |
| 52472M101 | LEGACY HOUSING CORP | 76022 | 2451 | TX |
| 45259A209 | IMPACT SHARE NCP MTY EMP ETF | 75034 | 6722 | TX |
| 75419T103 | RATTLER MDSTRM PRTNRS LP | 79701 | 4922 | TX |
| 205768302 | COMSTOCK RESOURCES INC | 75034 | 1311 | TX |
| 45259A100 | YWCA WOMENS EMPOWERMENT ETF | 75034 | 6722 | TX |
| 76118Y104 | RESIDEO TECHNOLOGIES | 85254 | 5065 | TX |
| 48253L205 | KLX ENERGY SERVS HLDNG | 77056 | 1389 | TX |
| 45259A308 | IMPACT SHRS SUTBL DV GAL ETF | 75034 | 6722 | TX |
| 02215L209 | KINETIK HOLDINGS INC | 79701 | 4922 | TX |
| 71948P100 | PHUNWARE INC | 78757 | 7370 | TX |
| 834251100 | SOLITON INC | 77081 | 3845 | TX |
| 87615L107 | TARGET HOSPITALITY CORP | 77381 | 6510 | TX |
| 10918L103 | BRIGHAM MINERALS INC | 78730 | 1311 | TX |
| 83946P107 | SOUTH PLAINS FINANCIAL INC | 79407 | 6020 | TX |
| 83946P107 | SOUTH PLAINS FINANCIAL INC | 79407 | 6020 | TX |
| 83548F200 | SONIM TECHNOLOGIES INC | 78730 | 3663 | TX |
| 86745K104 | SUNNOVA ENRGY INTL INC | 77046 | 4991 | TX |
| 14843C105 | CASTLE BIOSCIENCES INC | 77546 | 8071 | TX |
| 18978H102 | CNS PHARMACEUTICALS INC | 77027 | 2834 | TX |
| 87105M102 | SWITCHBACK ENERGY ACQ CORP | 75225 | 9995 | TX |
| 79400X107 | SALARIUS PHARMACEUTICALS INC | 77021 | 2836 | TX |
| 85236P101 | STABILIS SOLUTIONS INC | 77079 | 1311 | TX |
| 87241J104 | TFF PHARMACEUTICALS INC | 76107 | 2834 | TX |
| 64822P106 | NEW PROVIDENCE ACQ CORP | 78759 | 9995 | TX |
| 627333107 | MUSCLE MAKER INC | 77573 | 5812 | TX |
| 65342V101 | NEXPOINT REAL EST FIN | 75201 | 6798 | TX |
| 19424L101 | COLLECTIVE GROWTH CORP | 78701 | 9995 | TX |
| 29788T103 | E2OPEN PARENT HOLDINGS INC | 78759 | 7370 | TX |
| G8598Y109 | SUSTAINABLE OPP ACQ | 75201 | 9995 | TX |
| 51654W101 | LANTERN PHARMA INC | 75201 | 2836 | TX |
| 497634105 | KIROMIC BIOPHARMA INC | 77054 | 2836 | TX |
| 049430101 | ATLAS TECHNICAL CONSULTANTS | 78738 | 8700 | TX |
| 054748108 | AYRO INC | 78664 | 3711 | TX |
| 396879108 | GREENWICH LIFESCIE INC | 77477 | 2836 | TX |
| 68373J104 | OPEN LENDING CORP | 78746 | 7372 | TX |
| 750102105 | RACKSPACE TECHNOLO INC | 78218 | 7370 | TX |
| 92847W103 | VITAL FARMS INC | 78704 | 2015 | TX |
| 651448102 | NEWHOLD INVESTMENT COR | 77079 | 9995 | TX |
| 08975P108 | BIGCOMMERCE HOLDIN INC | 78726 | 7370 | TX |
| 64119V303 | NETSTREIT CORP | 75206 | 6798 | TX |
| 43114Q105 | HIGHPEAK ENERGY INC | 76102 | 1311 | TX |
| 877619106 | TAYSHA GENE THERPS INC | 75247 | 2836 | TX |
| G8990D125 | TPG PACE BENE FIN CORP | 76102 | 9995 | TX |
| G8990Y103 | TPG PACE TECH OPP CORP | 76102 | 9995 | TX |
| 51476H100 | LANDCADIA HOLD III INC | 77027 | 9995 | TX |
| 00402L107 | ACADEMY SPO AND OU INC | 77449 | 5940 | TX |
| 82024L103 | SHATTUCK LABS INC | 78703 | 2836 | TX |
| 886029206 | THRYV HOLDINGS INC | 75261 | 2741 | TX |
| 449109107 | HYLIION HOLDINGS CORP | 78613 | 3711 | TX |
| G1195N105 | BLUESCAPE OPP ACQ CORP | 75201 | 9995 | TX |
| 25434V104 | DIMENSIO US CORE EQ MARK ETF | 78746 | 6722 | TX |
| 25434V203 | DIMENSIO INTER CORE EQUI ETF | 78746 | 6722 | TX |
| 90470L550 | BALLAST SMALL/MID CAP ETF | 75219 | 6722 | TX |
| 25434V302 | DIMENSIONAL EM CR EQ MKT ETF | 78746 | 6722 | TX |
| 38113L107 | GOLDEN NUGGET ONLINE GAMING | 77027 | 7990 | TX |
| 12047B105 | BUMBLE INC | 78756 | 7370 | TX |
| 82671G100 | SIGNIFY HEALTH INC | 75244 | 8082 | TX |
| 01644J108 | ALKAMI TECHNOLOGY INC | 75024 | 7370 | TX |
| 45783C101 | INSTIL BIO INC | 75219 | 2836 | TX |
| 87652V109 | TASKUS INC | 78132 | 7374 | TX |
| 30320C103 | FTC SOLAR INC | 78759 | 3600 | TX |
| 034569103 | ANEBULO PHARMA INC | 78734 | 2834 | TX |
| 30322L101 | F45 TRAINING HLDNGS INC | 78704 | 7990 | TX |
| 126327105 | CS DISCO INC | 78746 | 7370 | TX |
| 29882P106 | EUROPEAN WAX CENTER INC | 75024 | 7200 | TX |
| 54911Q107 | LOYALTY VENTURES INC | 75024 | 7389 | TX |
| 238337109 | DAVE & BUSTER'S ENTMT INC | 75019 | 5810 | TX |
| 17299Z007 | CINEMARK USA INC | 75093 | 7830 | TX |
| 046484200 | ASTROTECH CORP | 78758 | 3829 | TX |
| 293306106 | ENGLOBAL CORP | 77079 | 8711 | TX |
| 30049A107 | EVOLUTION PETROLEUM CORP | 77079 | 1311 | TX |
| 721491108 | PILLARSTONE CAPITAL REIT | 77063 | 6512 | TX |
| 84470P109 | SOUTHSIDE BANCSHARES INC | 75701 | 6020 | TX |
| 84470P109 | SOUTHSIDE BANCSHARES INC | 75701 | 6020 | TX |
| 781846209 | RUSH ENTERPRISES INC | 78130 | 5500 | TX |
| 237545108 | DASSAULT SYSTEMS SA | 78140 | 7372 | |
| 268073103 | DYNARESOURCE INC | 75039 | 1040 | TX |
| 968235200 | WILHELMINA INTERNATIONAL INC | 75240 | 7900 | TX |
| 143905107 | CARRIAGE SERVICES INC | 77056 | 7200 | TX |
| 62955J103 | NOV INC | 77036-6565 | 3533 | TX |
| 371927104 | GENESIS ENERGY -LP | 77002 | 5172 | TX |
| 233377407 | DXP ENTERPRISES INC | 77040 | 5080 | TX |
| 45778Q107 | INSPERITY INC | 77339 | 7363 | TX |
| 402307102 | GULF ISLAND FABRICATION INC | 77084 | 3730 | TX |
| 199908104 | COMFORT SYSTEMS USA INC | 77057 | 1700 | TX |
| 42330P107 | HELIX ENERGY SOLUTIONS GROUP | 77043 | 1389 | TX |
| 03957W106 | ARCHROCK INC | 77024 | 3533 | TX |
| 023477300 | AMEN PROPERTIES INC | 75080 | 1311 | TX |
| 262037104 | DRIL-QUIP INC | 77041 | 3533 | TX |
| 68559A109 | ORBITAL INFRAS GROUP INC | 77038 | 1623 | TX |
| 398905109 | GROUP 1 AUTOMOTIVE INC | 77024 | 5500 | TX |
| 140475203 | SONIDA SENIOR LIVING INC | 75001 | 8300 | TX |
| 74762E102 | QUANTA SERVICES INC | 77056 | 1731 | TX |
| 20563P101 | COMPX INTERNATIONAL INC | 75240-2620 | 3420 | TX |
| 009119108 | AIR FRANCE - KLM | 75007 | 4512 | |
| 919134304 | VALEO SE | 75017 | 3714 | |
| 25375L206 | DIGERATI TECHNOLOGIES INC | 78216 | 4899 | TX |
| 659012108 | NORTH DALLAS BANK & TRUST CO | 75380-1826 | 6020 | TX |
| 659012108 | NORTH DALLAS BANK & TRUST CO | 75380-1826 | 6020 | TX |
| 57055L107 | MARKER THERAPEUTICS INC | 77027 | 2836 | TX |
| 926795204 | VIKING ENERGY GROUP | 77094 | 1311 | TX |
| 820017101 | SHARPS COMPLIANCE CORP | 77054 | 4955 | TX |
| 293792107 | ENTERPRISE PRODCT PARTNRS LP | 77002-5227 | 1311 | TX |
| 563771203 | MANNATECH INC | 75028 | 2834 | TX |
| 22822V101 | CROWN CASTLE INC | 77057-2261 | 6798 | TX |
| 88335R101 | THEGLOBE.COM INC | 75254 | 9995 | TX |
| 743606105 | PROSPERITY BANCSHARES INC | 77027 | 6020 | TX |
| 743606105 | PROSPERITY BANCSHARES INC | 77027 | 6020 | TX |
| 84856X106 | BITECH TECHNOLOGIES CORP | 92626 | 8090 | CA |
| 726503105 | PLAINS ALL AMER PIPELNE -LP | 77002 | 4610 | TX |
| 917313108 | USIO INC | 78231 | 6099 | TX |
| 095229100 | BLUCORA INC | 75019 | 6211 | TX |
| 23258P105 | CYNERGISTEK INC | 78759 | 7374 | TX |
| 23258P105 | CYNERGISTEK INC | 78759 | 7374 | TX |
| 444717102 | HUGOTON ROYALTY TRUST | 75219 | 6792 | TX |
| 87233Q108 | TC PIPELINES LP | 77002-2761 | 6799 | TX |
| 90333L201 | U S CONCRETE INC | 76039 | 3270 | TX |
| 794093104 | SALEM MEDIA GROUP INC | 75063 | 4832 | TX |
| 526107107 | LENNOX INTERNATIONAL INC | 75080 | 3585 | TX |
| 78501P203 | SWK HOLDINGS CORP | 75254 | 6159 | TX |
| 717098206 | PFSWEB INC | 75013 | 7374 | TX |
| 22161J206 | COSTAR TECHNOLOGIES INC | 75019 | 3651 | TX |
| 826919102 | SILICON LABORATORIES INC | 78701 | 3674 | TX |
| 55027E102 | LUMINEX CORP | 78727 | 2835 | TX |
| 528872302 | LEXICON PHARMACEUTICALS INC | 77381 | 2835 | TX |
| 762544104 | RIBBON COMMUNICATIONS INC | 75023 | 7373 | TX |
| 629377508 | NRG ENERGY INC | 77002 | 4911 | TX |
| 06985P209 | BASIC ENERGY SERVICES INC | 76102 | 1389 | TX |
| 14817C107 | CASSAVA SCIENCES INC | 78731 | 2834 | TX |
| 69372U207 | P10 HOLDINGS INC | 75205 | 6726 | TX |
| 04537Y109 | ASPIRA WOMENS HEALTH INC | 78738 | 2835 | TX |
| 45826H109 | INTEGER HOLDINGS CORP | 75024 | 3845 | TX |
| 67058H102 | NUSTAR ENERGY LP | 78257 | 4610 | TX |
| 678026105 | OIL STATES INTL INC | 77002 | 3533 | TX |
| 76129W105 | RETRACTABLE TECHNOLOGIES INC | 75068-5295 | 3841 | TX |
| 001744101 | AMN HEALTHCARE SERVICES INC | 75019 | 7361 | TX |
| 36467W109 | GAMESTOP CORP | 76051 | 5734 | TX |
| 15189W001 | CENTERPOINT ENERGY RES CORP | 77002 | 4923 | TX |
| 44183U209 | HOUSTON AMERN ENERGY CORP | 77002 | 1311 | TX |
| 262077100 | DRIVE SHACK INC | 75231 | 7990 | TX |
| 262077100 | DRIVE SHACK INC | 75231 | 7990 | TX |
| 390607109 | GREAT LAKES DREDGE & DOCK CP | 77024 | 1600 | TX |
| 44799X001 | HUNTSMAN INTERNATIONAL LLC | 77380 | 2800 | TX |
| 961765104 | WESTWOOD HOLDINGS GROUP INC | 75201 | 6282 | TX |
| 961765104 | WESTWOOD HOLDINGS GROUP INC | 75201 | 6282 | TX |
| 65336K103 | NEXSTAR MEDIA GROUP | 75062 | 4833 | TX |
| 573331105 | MARTIN MIDSTREAM PARTNERS LP | 75662 | 5172 | TX |
| 88224Q107 | TEXAS CAPITAL BANCSHARES INC | 75201 | 6020 | TX |
| 88224Q107 | TEXAS CAPITAL BANCSHARES INC | 75201 | 6020 | TX |
| 63886Q109 | NATURAL GAS SERVICES GROUP | 79705 | 7359 | TX |
| 63900P608 | NATURAL RESOURCE PARTNERS LP | 77002 | 6795 | TX |
| 044103869 | ASHFORD HOSPITALITY TRUST | 75254 | 6798 | TX |
| 044103869 | ASHFORD HOSPITALITY TRUST | 75254 | 6798 | TX |
| 15189X009 | CENTERPOINT ENRG HOUSTON ELE | 77002 | 4911 | TX |
| 68233D008 | ONCOR ELECTRIC DELIVERY CO | 75202 | 4911 | TX |
| 208242107 | CONN'S INC | 77381 | 5731 | TX |
| 432748101 | HILLTOP HOLDINGS INC | 75205 | 6199 | TX |
| 432748101 | HILLTOP HOLDINGS INC | 75205 | 6199 | TX |
| 50105F105 | KRONOS WORLDWIDE INC | 75240-2620 | 2810 | TX |
| 92840M102 | VISTRA CORP | 75039 | 4911 | TX |
| 84860W300 | SPIRIT REALTY CAPITAL INC | 75201 | 6798 | TX |
| 84860W300 | SPIRIT REALTY CAPITAL INC | 75201 | 6798 | TX |
| 84860W300 | SPIRIT REALTY CAPITAL INC | 75201 | 6798 | TX |
| 92922P106 | W&T OFFSHORE INC | 77057-5745 | 1311 | TX |
| 141743104 | CAREVIEW COMMUNICATIONS INC | 75067 | 7372 | TX |
| 960413102 | WESTLAKE CORP | 77056 | 2821 | TX |
| 253868103 | DIGITAL REALTY TRUST INC | 78735 | 6798 | TX |
| 253868103 | DIGITAL REALTY TRUST INC | 78735 | 6798 | TX |
| 024835100 | AMERICAN CAMPUS COMMUNITIES | 78738 | 6798 | TX |
| 024835100 | AMERICAN CAMPUS COMMUNITIES | 78738 | 6798 | TX |
| 486606106 | KAYNE ANDERSON MLP/MD INV CO | 77002 | 6726 | TX |
| 150870103 | CELANESE CORP | 75039-5421 | 2860 | TX |
| 98379L100 | XPEL INC | 78216 | 3714 | TX |
| 447011107 | HUNTSMAN CORP | 77380 | 2860 | TX |
| 87217L208 | TECTONIC FINANCIAL INC | 75248 | 6020 | TX |
| 87217L208 | TECTONIC FINANCIAL INC | 75248 | 6020 | TX |
| 218681104 | CORE MARK HOLDING CO INC | 76262 | 5190 | TX |
| 226344208 | CRESTWOOD EQUITY PARTNERS LP | 77002 | 5172 | TX |
| 12008R107 | BUILDERS FIRSTSOURCE | 75201 | 2452 | TX |
| 989696109 | ZION OIL & GAS INC | 75243 | 1311 | TX |
| 18453H106 | CLEAR CHANNEL OUTDOOR HLDGS | 78249 | 7310 | TX |
| 09057N300 | BIO-PATH HOLDINGS INC | 77401 | 2836 | TX |
| 29273V100 | ENERGY TRANSFER LP | 75225 | 4922 | TX |
| 916896103 | URANIUM ENERGY CORP | 78401 | 1090 | TX |
| 50077C106 | KRATON CORP | 77032-2348 | 2821 | TX |
| 25278X109 | DIAMONDBACK ENERGY INC | 79701 | 1311 | TX |
| 095395307 | BLUE DOLPHIN ENERGY CO | 77002 | 4922 | TX |
| 718546104 | PHILLIPS 66 | 77042 | 2911 | TX |
| 22944T109 | CUB ENERGY INC | T2P 2V7 | 5172 | AB |
| 34984V209 | FORUM ENERGY TECH INC | 77064 | 3533 | TX |
| 44244K109 | HOUSTON WIRE & CABLE CO | 77029 | 5063 | TX |
| 07134L107 | BATTALION OIL CORP | 77043 | 1311 | TX |
| 48242W106 | KBR INC | 77002 | 7370 | TX |
| 65340G205 | NEXPOINT DIVERSIFIED RE TS | 75201 | 6726 | TX |
| 30053M104 | EVOLVE TRANSITION INFRAST LP | 77056 | 4610 | TX |
| 74164F103 | PRIMORIS SERVICES CORP | 75201 | 1623 | TX |
| 79546E104 | SALLY BEAUTY HOLDINGS INC | 76210 | 5990 | TX |
| 05366Y201 | AVIAT NETWORKS INC | 78728 | 3663 | TX |
| 76680V108 | RING ENERGY INC | 77380 | 1311 | TX |
| 16411Q101 | CHENIERE ENERGY PARTNERS LP | 77002 | 5171 | TX |
| 17243V102 | CINEMARK HOLDINGS INC | 75093 | 7830 | TX |
| 87288V101 | TSS INC | 78664 | 8741 | TX |
| 74346Y103 | PROS HOLDINGS INC | 77098 | 7370 | TX |
| 231631300 | CUSHING MLP & INFASTCR TR FD | 75201 | 6726 | TX |
| 56035L104 | MAIN STREET CAPITAL CORP | 77056 | 6797 | TX |
| 56035L104 | MAIN STREET CAPITAL CORP | 77056 | 6797 | TX |
| 12662P108 | CVR ENERGY INC | 77479 | 2911 | TX |
| 294375209 | EPSILON ENERGY LTD | 77060 | 1311 | TX |
| 294375209 | EPSILON ENERGY LTD | 77060 | 1311 | TX |
| 67091K302 | NUVERRA ENVIRONMENTAL SOLUTN | 77079 | 1389 | TX |
| 68628V308 | ORION GROUP HOLDINGS INC | 77034 | 1600 | TX |
| 29399W008 | ENTERGY TEXAS INC | 77380 | 4911 | TX |
| 235050101 | DALLASNEWS CORP | 75201-4866 | 2711 | TX |
| 83417Q204 | SOLARWINDS CORP | 78735 | 7370 | TX |
| 55345K103 | MRC GLOBAL INC | 77010 | 5051 | TX |
| 39679T104 | GREENWAY TECHNOLOGIES INC | 76011 | 3533 | TX |
| 12637A103 | CSI COMPRESSCO LP | 77380 | 1382 | TX |
| 55314K506 | MMEX RESOURCES CORP | 78735 | 2911 | TX |
| 90346E103 | U S SILICA HOLDINGS INC | 77494 | 1400 | TX |
| 25470P102 | DISCOVERY ENERGY CORP | 77056 | 1311 | TX |
| 006739106 | ADDUS HOMECARE CORP | 75034 | 8082 | TX |
| 516808102 | LAREDO OIL INC | 78705 | 1382 | TX |
| 92534K107 | VERTEX ENERGY INC | 77058 | 5172 | TX |
| 674215207 | CHORD ENERGY CORP | 77002 | 1311 | TX |
| 75606N109 | REALPAGE INC | 75082-4305 | 7370 | TX |
| 84612H106 | SOW GOOD INC | 75061 | 2030 | TX |
| 866082100 | SUMMIT HOTEL PROPERTIES INC | 78738 | 6798 | TX |
| 866082100 | SUMMIT HOTEL PROPERTIES INC | 78738 | 6798 | TX |
| 966084204 | WHITESTONE REIT | 77063 | 6798 | TX |
| 966084204 | WHITESTONE REIT | 77063 | 6798 | TX |
| 87612G101 | TARGA RESOURCES CORP | 77002 | 4923 | TX |
| 44267D107 | HOWARD HUGHES CORP | 77380 | 6552 | TX |
| 766582100 | RIGNET INC | 77084-4947 | 4899 | TX |
| 48661E108 | KAYNE ANDERSON MDSTRM/ENG FD | 77002 | 6726 | TX |
| 89102U103 | TORCHLIGHT ENERGY RESOURCES | 75093 | 1311 | TX |
| 21077F100 | CONTANGO ORE INC | 77098 | 1040 | TX |
| 126633205 | CVR PARTNERS LP | 77479 | 2870 | TX |
| 91829B103 | VOC ENERGY TRUST | 77002 | 1311 | TX |
| 500688106 | KOSMOS ENERGY LTD | 75231 | 1311 | TX |
| 88362T103 | THERMON GROUP HOLDINGS INC | 78735 | 3690 | TX |
| 82938H107 | SINTANA ENERGY INC | 75225 | 1311 | TX |
| 59833H101 | MIDWEST ENERGY EMISSIONS CP | 75109 | 3590 | TX |
| 171604101 | CHUY'S HOLDINGS INC | 78704 | 5812 | TX |
| 576485205 | MATADOR RESOURCES CO | 75240 | 1311 | TX |
| 31660B101 | FIESTA RESTAURANT GROUP INC | 75254 | 5812 | TX |
| 03212B103 | AMPLIFY ENERGY CORP | 77002 | 1311 | TX |
| 684060106 | ORANGE | 92130 | 4813 | |
| 54240F202 | LONESTAR RESOURCES US INC | 76107 | 1311 | TX |
| 72941H400 | PLUS THERAPEUTICS INC | 78756 | 2836 | TX |
| 12504L109 | CBRE GROUP INC | 75201 | 6500 | TX |
| 12504L109 | CBRE GROUP INC | 75201 | 6500 | TX |
| G1991C105 | CARDTRONICS PLC | 77042 | 7389 | TX |
| 435763107 | HOLLY ENERGY PARTNERS LP | 75201-1507 | 4610 | TX |
| 15117K103 | CELLECTIS SA | 75013 | 2836 | |
| N53745100 | LYONDELLBASELL INDUSTRIES NV | 77010 | 2820 | TX |
| 039653100 | ARCOSA INC | 75201 | 3440 | TX |

**Table 2: Descriptive Statistics for power outage affected companies**

This Table shows the descriptive statistics of the sample of power outage affected companies for year 2021. Firm size is the natural logarithm of total assets. Market value is the market value of the firm in million USD. Profitability of the firm is the return on assets defined as EBITDA scaled by total assets. Leverage is the sum of short-term and long-term debt scaled by book assets.

| **Variable** | **Number of observations** | **Mean** | **Standard deviation** | **Min** | **Median** | **Max** |
|---|---|---|---|---|---|---|
| Firm size | 550 | 6.862 | 2.843 | -6.908 | 7.052 | 14.91 |
| Leverage | 320 | -1.68 | 0.067 | 35.43 | -129 | 20.14 |
| Market value of the firm | 509 | 5,191 | 19,302 | 0.708 | 560.6 | 221,574 |
| Profitability of the firm | 480 | -2.97 | 59.64 | -1302 | 0.044 | 30.51 |

**Figure 1: The effect of the power outage on the actual return of firms by using the market adjustment model**

This figure shows the CAR by using four different models. The CAR are calculated based on the market adjustment model, market model, three factor model, and momentum model in Panels A, B, C, and D. We discuss the setup of the event model in Section 4. Empirical results. The full line is the evolution of the mean CAR. The dashed lines are 95% confidence intervals.

Panel A: Market adjustment model

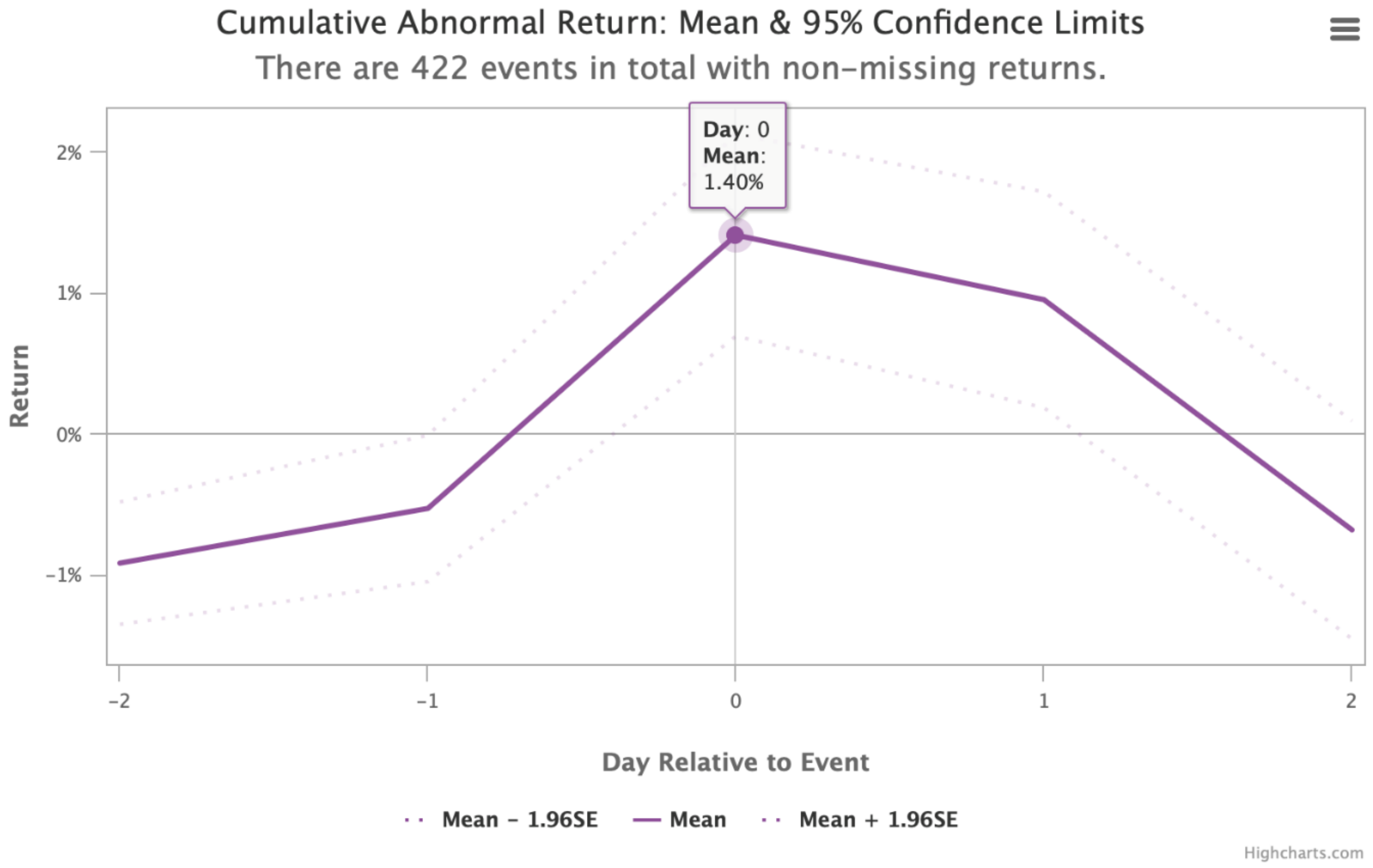

Panel B: Market model

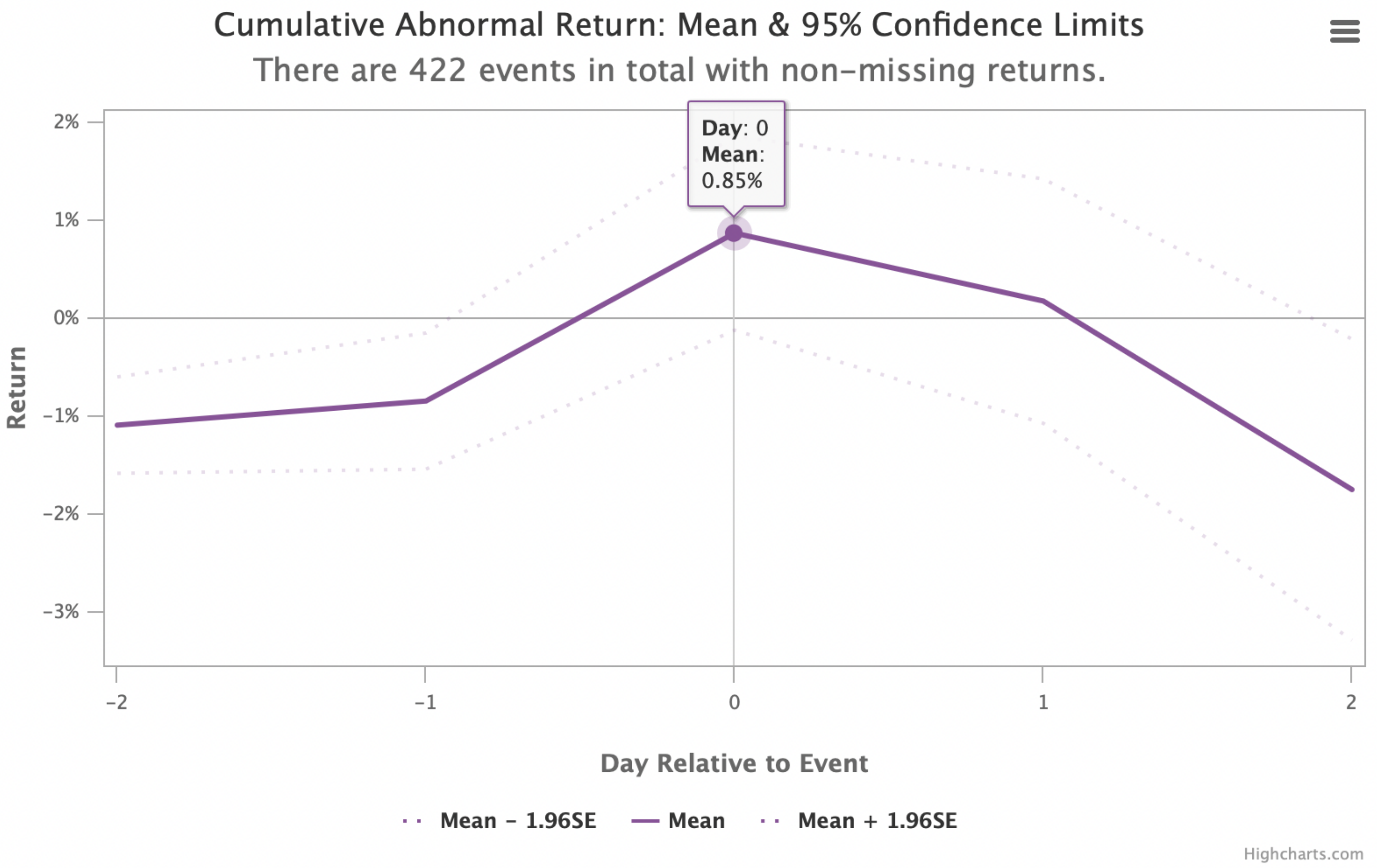

Panel C: Three factor model

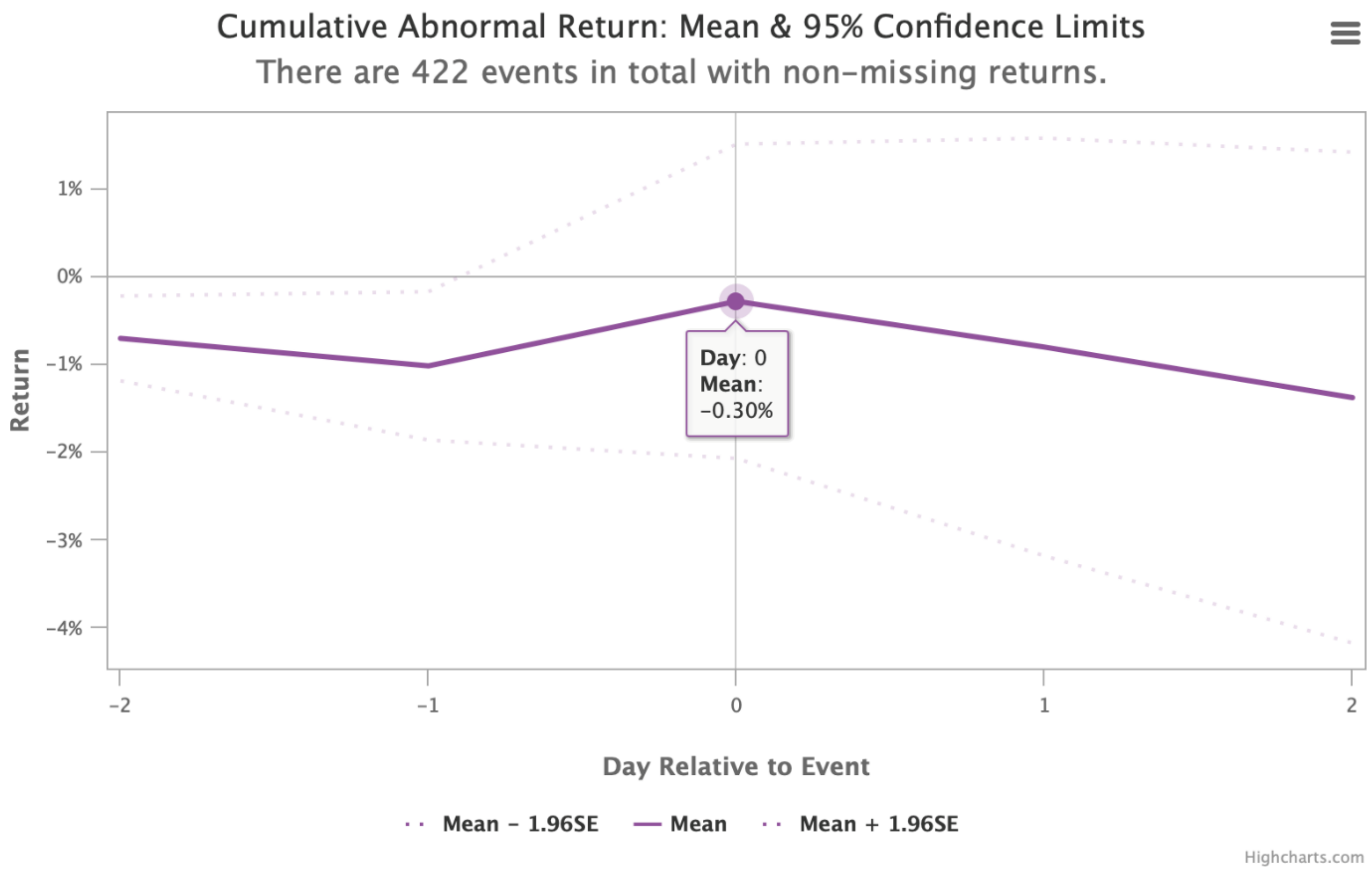

Panel D: Momentum model

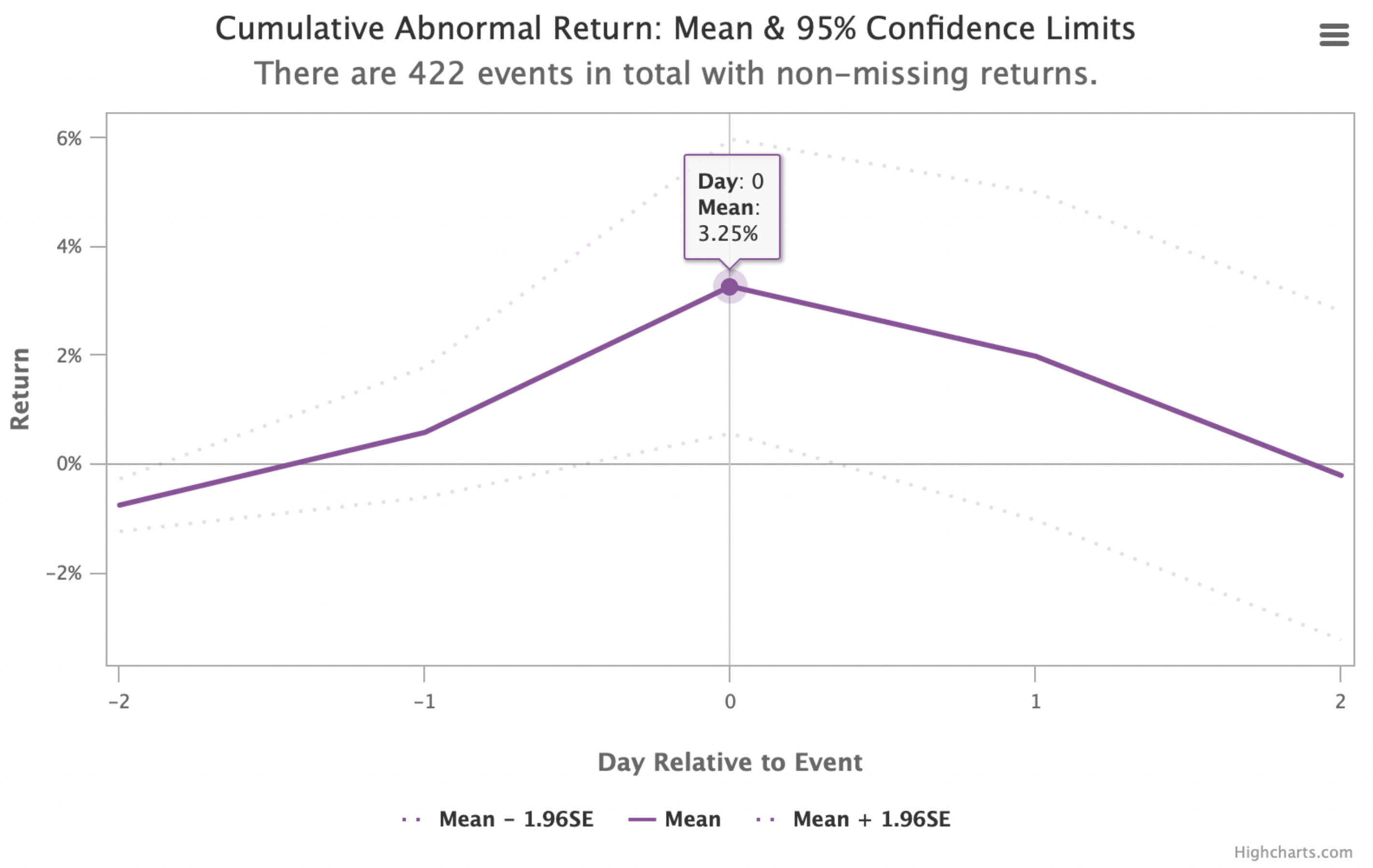